\documentclass[useAMS,usenatbib]{mn2e}
\pdfoutput=1
\usepackage{amsmath}
\usepackage{txfonts}
\usepackage{graphicx}
\usepackage{xspace}
\usepackage{flushend}
\usepackage[colorlinks,linkcolor=blue,citecolor=blue,urlcolor=blue,pdftitle={Weak gravitational lensing with DEIMOS},pdfauthor={Peter Melchior et al.}]{hyperref}

\newcommand{\mom}[2]{%
\lbrace #1 \rbrace_{#2}%
}
\newcommand{\deimos}{%
{\scshape Deimos}\xspace%
}

\title[Weak gravitational lensing with DEIMOS]{Weak gravitational lensing with {\scshape Deimos}}
\author[P. Melchior et al.]{P. Melchior$^{1}$\thanks{E-mail:
pmelchior@ita.uni-heidelberg.de}, M. Viola$^{1}$, B. M. Sch\"{a}fer$^{2}$, and M. Bartelmann$^{1}$\\
$^{1}$Zentrum f\"ur Astronomie Heidelberg, Institut f\"ur Theoretische Astrophysik, Albert-Ueberle-Str. 2, D-69120 Heidelberg, Germany\\
$^{2}$Zentrum f\"ur Astronomie Heidelberg, Astronomisches Rechen-Institut, M\"{o}nchhofstr. 12-14, D-69120 Heidelberg, Germany}

\begin{document}

\date{Accepted 2010 October 14. Received 2010 October 8; in original form 2010 August 9}

\pagerange{\pageref{firstpage}--\pageref{lastpage}} \pubyear{2010}

\maketitle
\label{firstpage}

\begin{abstract}
We introduce a novel method for weak-lensing measurements, which is
based on a mathematically exact deconvolution of the moments of the
apparent brightness distribution of galaxies from the telescope's
PSF. No assumptions on the shape of the galaxy or the PSF are
made. The (de)convolution equations are exact for unweighted
moments only, while in practice a compact weight function needs to
be applied to the noisy images to ensure that the moment measurement
yields significant results. We employ a Gaussian weight function,
whose centroid and ellipticity are iteratively adjusted to match the
corresponding quantities of the source. The change of the moments
caused by the application of the weight function can then be
corrected by considering higher-order weighted moments of the same
source. Because of the form of the deconvolution equations, even an
incomplete weighting correction leads to an excellent shear estimation
if galaxies and PSF are measured with a weight function of
identical size. 

We demonstrate the accuracy and capabilities of this new method in the
context of weak gravitational lensing measurements with a set of
specialized tests and show its competitive performance on the GREAT08
challenge data. A complete C++ implementation of the method can be requested from the authors.
\end{abstract}


\section{Introduction}

Shear estimation from noisy galaxy images is a challenging
task, even more so for the stringent accuracy requirements of upcoming cosmic
shear surveys. Existing methods can achieve multiplicative errors in the percent range \citep{Bridle10.1}, but to exploit the statistical power of the next-generation surveys errors in the permille range or even below are requested \citep{Amara08.1}.

Shear estimates can be achieved in a model-based or in a
model-independent fashion. For instance, {\scshape Lensfit}
\citep{Miller07.1} compares sheared and convolved bulge-disk profiles to the given galaxies. Model-based approaches often perform excellently for strongly degraded data because certain implicit or explicit priors keep the results within reasonable bounds, e.g. the source ellipticity smaller than unity. On the other hand, when imposing these priors to data, whose characteristics differ from the expectation, these approaches may also bias the outcome. 

Model-independent approaches do not -- or at least not as strongly --
assume particular knowledge of the data to be analyzed. They should
therefore generalize better in applications, where priors are not
obvious, e.g. on the intrinsic shape of lensed galaxies. The
traditional KSB method \citep{Kaiser95.1} forms a shear estimator from
the second-order moments of lensed galaxy images. When doing so, it is
not guaranteed that reasonable shear estimates can be achieved for
each galaxy. Consequently, KSB requires a careful setup, which is
adjusted to the characteristics of the data to be analyzed. KSB
furthermore employs strong assumptions on the PSF shape, which are not
necessarily fulfilled for a given telescope or observation
\citep{Kuijken99.1}. As we have shown recently, KSB relies on several other assumptions concerning the relation between convolved and unconvolved ellipticity as well as the relation between ellipticity and shear, neither of which hold in practice \citep{Viola10.1}.

In this work we present a novel method for shear estimation, which maintains the strengths of model-independent approaches by working with multipole moments, but does not suffer from the KSB-shortcomings mentioned above.

\section[The DEIMOS method]{The D{\sevensize\bf EIMOS} method}
\label{sec:deimos}
The effect of gravitational lensing on the surface brightness distribution $G(\mathbf{x})$ of a distant background galaxies is most naturally described in terms of the moments of the brightness distribution,
\begin{equation}
\label{eq:2d_moment}
\mom{G}{i,j} \equiv \int d^2x\ G(\mathbf{x})\, x_1^i x_2^j,
\end{equation}
for which we introduce a tensor-like notation. The effect of the reduced shear $\mathbf{g}$ is contained in the change of the complex ellipticity
\begin{equation}
\label{eq:ellipticity}
 \bchi \equiv \frac{\mom{G}{2,0} - \mom{G}{0,2} + 2\mathrm{i}\mom{G}{1,1}}{\mom{G}{2,0} + \mom{G}{0,2}}
\end{equation}
with respect to its value before lensing,
\begin{equation}
\label{eq:lensing_chi}
  \bchi^s = \frac{\bchi - 2\mathbf{g} + \mathbf{g}^2\bchi^\star}{1+|\mathbf{g}|^2-2\Re(\mathbf{g}\bchi^\star)}
\end{equation}
 \citep[e.g.][]{Bartelmann01.1}. Unfortunately, we do not know $\bchi^s$, which would allow solving directly for $\mathbf{g}$ given $\bchi$. Furthermore, (at
least) two observational complications alter the source's moments,
typically much more drastically than lensing: convolution with the PSF
and any means of noise reduction -- normally weighting with a compact
function -- to 
yield significant moment measurements. Hence, $\bchi$ is not directly
accessible and needs to be estimated by properly accounting for the
observation effects. We leave the treatment of weighting for \autoref{sec:weighting} and start with the derivation of the change of the moments under convolution.

Any square-integrable one-dimensional function $G(x)$, has an exact representation in Fourier space,
\begin{equation}
\label{eq:char_function}
 G(x) \rightarrow \phi_G(k) = \int dx\, G(x)\, \mathrm{e}^{\mathrm{i}kx}.
\end{equation}
In the field of statistics, $\phi_G$ is often called the \emph{characteristic function} of $G$ and has a notable alternative form\footnote{The summation indices in this work all start with zero unless explicitly noted otherwise.}
\begin{equation}
\label{eq:char_function_moments}
 \phi_G(k) = \sum_n^\infty \mom{G}{n} \frac{(\mathrm{i}k)^n}{n!},
\end{equation}
which provides a link between the Fourier-transform of $G$ and its moments $\mom{G}{n}$, the one-dimensional pendants to \autoref{eq:2d_moment}.
We can now employ the convolution theorem, which allows us to replace the convolution by a product in Fourier-space, i.e. by a product of characteristic functions of $G$ and of the PSF kernel $P$,
\begin{equation}
 G^\star(x) \equiv \int dx'\, G(x)\, P(x-x') \rightarrow \phi_{G^\star} = \phi_G\, \phi_P.
\end{equation}
For convenience, we assume throughout this work the PSF to be flux-normalized, $\mom{P}{0}=1$.
Considering \autoref{eq:char_function_moments}, we get
\begin{equation}
 \begin{split}
 \phi_{G^\star}(k) &= \Bigl[\sum_n^\infty \mom{G}{n}\frac{(\mathrm{i}k)^n}{n!}\Bigr]
  \Bigl[\sum_n^\infty \mom{P}{n}\frac{(\mathrm{i}k)^n}{n!}\Bigr]\\
 &= \sum_n^\infty \sum_m^n \mom{G}{m}\frac{(\mathrm{i}k)^m}{m!} \mom{P}{n-m}\frac{(\mathrm{i}k)^{n-m}}{(n-m)!}\\
& = \sum_n^\infty\Bigl[\sum_m^n \tbinom{n}{m}\mom{G}{m} \mom{P}{n-m}\Bigr]\frac{(\mathrm{i}k)^n}{n!},
 \end{split}
\end{equation}
where we applied the Cauchy product in the second step. The expression in square brackets on the last line is by definition the desired moment,
\begin{equation}
\mom{G^\star}{n} = \sum_m^n \tbinom{n}{m}\mom{G}{m} \mom{P}{n-m}.
\end{equation}
 Hence, we can now express a convolution of the function $G$ with the kernel $P$ entirely in moment space. Moreover, even though the series in \autoref{eq:char_function_moments} is infinite, the order of the moments occurring in the computation of $\mom{G^\star}{n}$ is bound by $n$. This means, for calculating all moments of $G^\star$ up to order $n$, the knowledge of the same set of moments of $P$ and $G$ is completely sufficient. This results hold for any shape of $G$ and $P$ as long as their moments do not diverge. For non-pathological distributions, this requirement does not pose a significant limitation.

An identical derivation can be performed for two-dimensional
moments, which yields the change of moments of $G(\mathbf{x})$ under
convolution with the kernel $P(\mathbf{x})$ \citep{Flusser98.1}:

\begin{equation}
\label{eq:2d_convolution_moment}
 \mom{G^\star}{i,j} = \sum_k^{i} \sum_l^{j} \tbinom{i}{k} \tbinom{j}{l} \mom{G}{k,l} \mom{P}{i-k,j-l}\,.
\end{equation}

\subsection*{Deconvolution}
To obtain the deconvolved moments required for the shear estimation
via the ellipticity $\bchi$, we need to measure the moments up to
second order of the convolved galaxy shape and of the PSF kernel
shape. Then we can make use of a remarkable feature of
\autoref{eq:2d_convolution_moment}, which is already apparent from its
form: The impact of convolution on a moment of order $i+j=n$ is only a
function of unconvolved moments of lower order and PSF moments of at
most the same order. We can therefore start in zeroth order, the 
flux, which only needs to be corrected if the PSF is not flux-normalized. With the accurate value of the zeroth order and the first
moments of the PSF, we can correct the first-order moments of the
galaxy, and so on (cf.\autoref{tab:2nd_moments}). The hierarchical build-up of the deconvolved moments is the heart of the \deimos method (short for \emph{deconvolution in moment space}).

It is important to note and will turn out to be crucial for weak-lensing applications that with this deconvolution scheme we do not need to explicitly address the pixel noise, which hampers most other deconvolution approaches in the frequency domain, simply because we restrict ourselves to inferring the most robust low-order moments only.

\begin{table}
\caption{Equations for deconvolving all moments up to order $n=2$. The
  shown equations are specializations of \autoref{eq:2d_convolution_moment}.}
\label{tab:2nd_moments}
\begin{tabular}{l}
\hline
$\mom{G}{0,0}\ \mom{P}{0,0} = \mom{G^\star}{0,0}$\\[\smallskipamount]
$\mom{G}{0,1}\ \mom{P}{0,0} = \mom{G^\star}{0,1} - \mom{G}{0,0}\ \mom{P}{0,1}$\\[\smallskipamount]
$\mom{G}{1,0}\ \mom{P}{0,0} = \mom{G^\star}{1,0} - \mom{G}{0,0}\ \mom{P}{1,0}$\\[\smallskipamount]
$\mom{G}{0,2}\ \mom{P}{0,0} = \mom{G^\star}{0,2} - \mom{G}{0,0}\ \mom{P}{0,2} - 2\mom{G}{0,1}\ \mom{P}{0,1}$\\[\smallskipamount]
$\mom{G}{1,1}\ \mom{P}{0,0} = \mom{G^\star}{1,1} - \mom{G}{0,0}\ \mom{P}{1,1} - \mom{G}{0,1}\ \mom{P}{1,0} - \mom{G}{1,0}\ \mom{P}{0,1}$\\[\smallskipamount]
$\mom{G}{2,0}\ \mom{P}{0,0} = \mom{G^\star}{2,0} - \mom{G}{0,0}\ \mom{P}{2,0} - 2\mom{G}{1,0}\ \mom{P}{1,0}$\\[\smallskipamount]

\hline
\end{tabular}
\end{table}

\section{Noise and weighting}
\label{sec:weighting}

In practice, the moments are measured from noisy image data,
\begin{equation}
 I(\mathbf{x}) = G(\mathbf{x}) + N(\mathbf{x}),
\end{equation}
where the noise $N$ can be considered to be independently drawn from a Gaussian distribution with variance $\sigma_n^2$, i.e. $\langle N(\mathbf{x}_i) N(\mathbf{x}_j) \rangle = \sigma_n^2 \delta_{ij}$ for any two positions $\mathbf{x}_i$ and $\mathbf{x}_j$. According to \autoref{eq:2d_moment}, the image values at large distances from the galactic center have the largest impact on the $\langle I \rangle_n$ if $n>0$. For finite and compact brightness distributions $G$, these values are dominated by the noise process instead of the galaxy, whose moments we seek to measure. Consequently, centered weight functions $W$ of finite width are typically introduced to limit the integration range in \autoref{eq:2d_moment} to regions in which $I$ is mostly determined by $G$,
\begin{equation}
\label{eq:weighting}
 I_w(\mathbf{x}) \equiv W(\mathbf{x})\, I(\mathbf{x}).
\end{equation}
A typical choice for $W$ is a circular Gaussian centered at the galactic centroid,
\begin{equation}
\label{eq:Gaussian}
 W(\mathbf{x}) \equiv \exp\Bigl(-\frac{\mathbf{x}^2}{2s^2}\Bigr).
\end{equation}
Alternatively, one can choose to optimize the weight function to the shape of the source to be measured. \citet[see their section 3.1.2]{Bernstein02.1} proposed the usage of a Gaussian, whose centroid $\mathbf{x}_c$, size $s$, and ellipticity $\bepsilon$ are matched to the source, such that the argument of the exponential in \autoref{eq:Gaussian} is modified according to
\begin{equation}
\label{eq:eps_trafo}
 \mathbf{x} \to \mathbf{x}^\prime = \begin{pmatrix}
1-\epsilon_1 & -\epsilon_2\\
-\epsilon_2 & 1+\epsilon_1
\end{pmatrix}
(\mathbf{x}-\mathbf{x}_c).
\end{equation}
As such a weight function represents a matched spatial filter, it optimizes the significance and accuracy of the measurement if its parameters are close to their true values. This can, however, not be guaranteed in presence of pixel noise, but we found the iterative algorithm proposed by \citet{Bernstein02.1} to converge well in practice and therefore employ it to set the weight function within the \deimos method.

Unfortunately, a product in real space like the one in
\autoref{eq:weighting} translates into a convolution in
Fourier-space. We therefore have to expect some amount of mixing of
the moments of $I_w$. Even worse, an attempt to relate the moments of
$I_w$ to those of $I$ leads to diverging integrals. Hence, there is
no exact way of incorporating spatial weighting to the moment approach
outlined above. On the other hand, we can invert
\autoref{eq:weighting} for $I=I_w/W$ and expand $1/W$ in a Taylor
series around the center at $\mathbf{x}=\mathbf{0}$,
\begin{equation}
\begin{split}
 W^{-1}(\mathbf{x})\ \approx\ &W^{-1}(\mathbf{0}) - W'(\mathbf{0})\Bigl[\sum_{k=1}^2 c_k x_k^2\ + 4 \epsilon_2 x_1 x_2\Bigr]\ +\\
&\frac{1}{2} W''(\mathbf{0})\Bigl[\sum_{k,l=1}^2 c_k c_l x_k^2 x_l^2 - 8 \epsilon_2 \sum_{k=1}c_k x_k^2 x_1 x_2 + (4\epsilon_2 x_1 x_2)^2\Bigr],
\end{split}
\end{equation}
where we employed $W'(\mathbf{x})\equiv \frac{dW(\mathbf{x})}{d\mathbf{x}^2}$ and $c_{1,2} \equiv (1\mp\epsilon_1)^2 + \epsilon_2^2$. We introduce the parameter $n_w$ as the maximum order of the Taylor expansion, here $n_w=4$. Inserting this expansion in \autoref{eq:2d_moment}, we are able to
approximate the moments of $I$ by their \emph{deweighted} counterparts $\mom{I_{dw}}{}$. For convenience we give the correction terms for orders $n_w \leq 6$ in \autoref{tab:deweighting}.
This linear expansion allows us to correct for the weighting-induced
change in the moments of a certain order $n$ by considering the impact of the weight function on weighted moments up to order $n+n_w$.

\begin{table}
\caption{Correction terms for deweighting moments of order $n=i+j$. The deweighted moments $\mom{I_{dw}}{i,j}$ are given by the sum of the correction terms up to the limiting order $n_w$.}
\label{tab:deweighting}
\begin{tabular}{ll}
$n_w$ & correction terms\\
\hline
0 & $\mom{I_w}{i,j}$\\[\medskipamount]
2 & $\frac{1}{2s^2}\Bigl[c_1\mom{I_w}{i+2,j} - 4\,\epsilon_2 \mom{I_w}{i+1,j+1} + c_2 \mom{I_w}{i.j+2}\Bigr]$\\[\medskipamount]
4 & $\frac{1}{8s^4}\Bigl[ c_1^2 \mom{I_w}{i+4,j} - 8\,c_1 \epsilon_2 \mom{I_w}{i+3,j+1} + \bigl[ 2\,c_1 c_2 + 16\,\epsilon_2^2\bigr],\mom{I_w}{i+2,j+2}\ -$\\[\smallskipamount]
& $\phantom{\frac{1}{8s^4}\Bigl[}8\,c_2 \epsilon_2 \mom{I_w}{i+1,j+3} + c_2^2 \mom{I_w}{i,j+4}\Bigr]$\\[\medskipamount]
6 & $\frac{1}{48s^6}\Bigl[ c_1^3 \mom{I_w}{i+6,j} - 12\,c_1^2
\epsilon_2 \mom{I_w}{i+5,j+1} + \bigl[ 3\,c_1^2 c_2 + 48\,c_1
\epsilon_2^2\bigr]\, \mom{I_w}{i+4,j+2}\ -$\\[\smallskipamount]
& $\phantom{\frac{1}{48s^6}\Bigl[}\bigl[ 24\,c_1 c_2 \epsilon_2 + 64\,\epsilon_2^3\bigr]\, \mom{I_w}{i+3,j+3}\ + \, \bigl[ 3\,c_1 c_2^2 + 48\,c_2 \epsilon_2^2\bigr]\, \mom{I_w}{i+2,j+4}\ -$\\[\smallskipamount]
 & $\phantom{\frac{1}{48s^6}\Bigl[} 12\,c_2^2 \epsilon_2 \mom{I_w}{i+1,j+5} + c_2^3 \mom{I_w}{i,j+6}\Bigr]$\\
\hline
\end{tabular}
\end{table}

\subsection{Deweighting bias}
\label{sec:deweighting_bias}
The truncation of the Taylor expansion constitutes
the first and only source of bias in the \deimos
method. The direction of the bias is evident: As the weight function
suppresses contributions to the moments from pixel far away from the
centroid, its employment reduces the power in any moment by an amount,
which depends on the shape -- particularly on the radial profile -- of
the source and the width $s$. Additionally, if the ellipticity
$\bepsilon$ was misestimated during the matching of $W$, the measured
ellipticity of the source $\bchi$ before and after deweighting will be
biased towards $\bepsilon$. In the realistic case of noisy
images, which we address in more detail in \autoref{sec:deweighting_variance},
$\bepsilon$ can be wrong in two ways: statistically and
systematically. The centered Gaussian distribution of the pixel noise leads to
a centered Cauchy-type error distribution of both components of
$\bepsilon$, i.e. the statistical errors of $\bepsilon$ and therefore
also $\chi$ have vanishing mean. The systematic errors stem from the
application of a compact weight function to measure $\epsilon$, which for
any finite $s$ constitutes a removal of information. This leads to
deviations of the measured $\bepsilon$ from the true $\bepsilon$ if
e.g. the centroid is not determined accurately or the ellipticity
changes with radius. For individual objects, these deviations are
impossible to quantify precisely -- as this would require knowledge of
the true observed morphology -- and hamper all weak-lensing
measurements unless an appropriate treatment is devised
\citep[e.g.][]{Hosseini09.1, Bernstein10.1}. For large ensembles of
galaxies, effective levels of systematic errors can be assessed by
analyzing dedicated simulations (cf. \autoref{sec:great08}).

We investigate now the \deimos -specific systematic impact of a finite $n_w$ on the recovery of the deweighted moments. For the experiments in this section we simulated simple galaxy models following the S{\'e}rsic profile
\begin{equation}
 \label{eq:sersic_profile}
 p_s(r) \propto \exp\Bigl\lbrace -b_{n_s}\Bigl[\Bigl(\frac{r}{R_e}\Bigr)^{1/n_s}-1\Bigr]\Bigr\rbrace, 
\end{equation}
where $n_s$ denotes the S{\'e}rsic index, and $R_e$ the effective radius, while the PSFs are modeled from the Moffat profile
\begin{equation}
\label{eq:moffat_profile}
p_m(r) \propto (1+\alpha r^2)^{-\beta},
\end{equation}
where $\alpha =
\bigl(2^{1/\beta}-1\bigl)/\bigl(\mathrm{FWHM}/2\bigr)^2$ sets the
width of the profile and $\beta$ its slope. Both model types acquire their ellipticity according to \autoref{eq:eps_trafo}

\begin{figure}
\includegraphics[width=\linewidth]{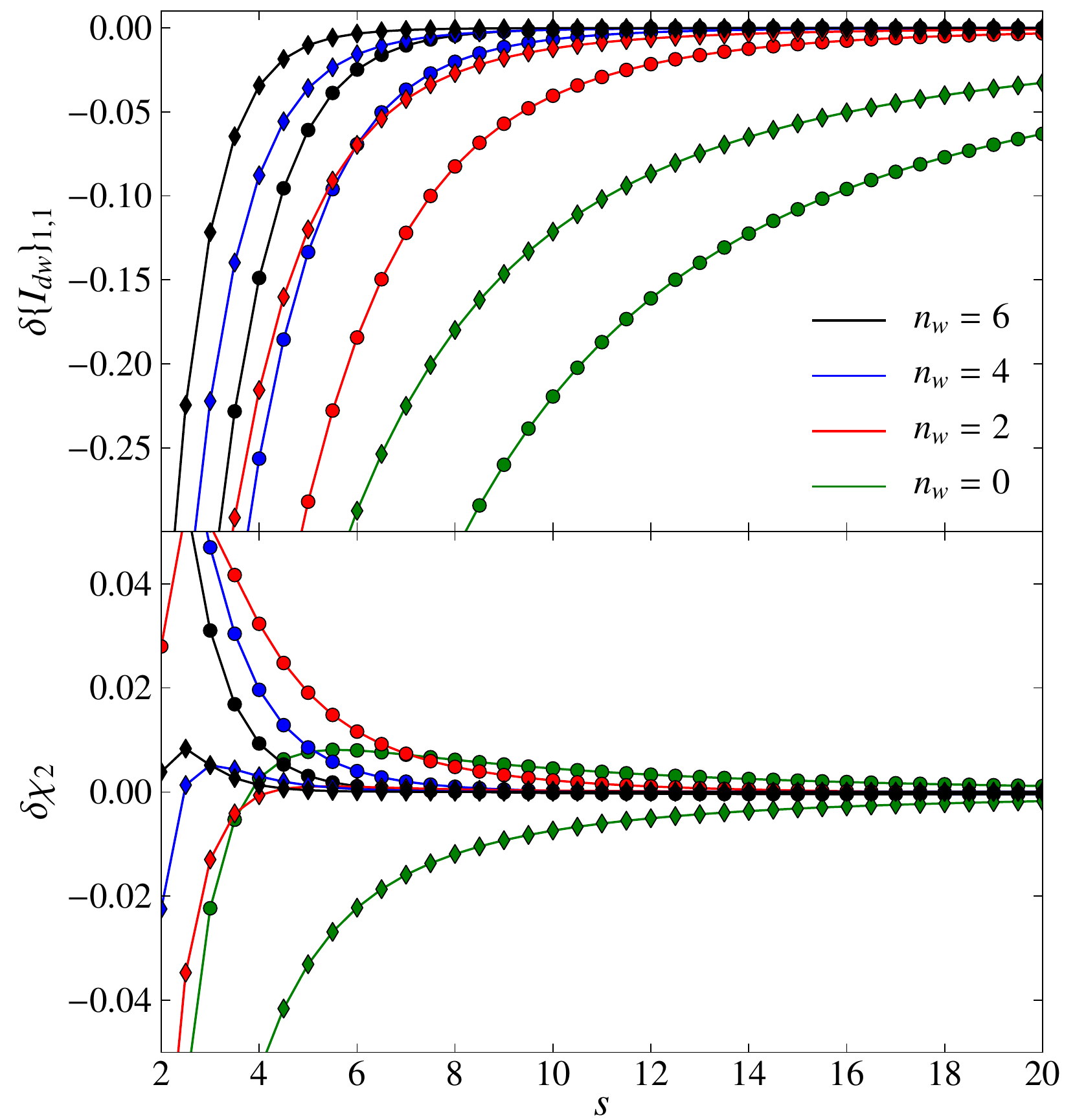}
\caption{Weighting-induced bias. Relative error of the PSF-convolved
  and deweighted moment $\mom{I_{dw}}{1,1}$ (top) and of the estimated ellipticity after deconvolution (bottom) as a function of weight function size $s$. Colors indicate the correction order $n_w$, while markers denote the galaxy model (circles for model 1, diamonds for model 2). The galaxy models are simulated as S{\'e}rsic profiles with the following parameters: $\bepsilon^s = (0.1,0.4),\ n_s = 1\,(4),\ R_e = 3\,(1.5)$ for model 1 (2). The PSF was of Moffat-type with $\bepsilon = (0.05,0.05),\ \beta=3,\ \mathrm{FWHM}=3$. The weight functions of galaxy and PSF had the same size $s$, but individually matched ellipticities.}
\label{fig:bias_weighting}
\end{figure}

In the top panel of \autoref{fig:bias_weighting} we show the error after deweighting a convolved galaxy image from a matched elliptical weight function as a function of its size $s$. As noted above, the bias is always negative and is clearly more prominent for the larger disk-type galaxy (circle markers). As the Taylor expansion becomes more accurate for $n_w\to\infty$ or $s\to\infty$, the bias of any moment decreases accordingly. 

An important consequence of the employment of a weight function
with matched ellipticity is that the bias after deweighting does only
very weakly depend on the apparent ellipticity, i.e. all moments of
the same order are biased by the same relative factor
$\Delta(n,s)$. This means any ratio of such moments remains
unbiased. This does not guarantee that the ellipticity is still
unbiased after the moments have passed the deconvolution step, which
is exact only for unweighted moments. On the other hand, the
particular form of the equations in \autoref{tab:2nd_moments} becomes
important here: If we assume well-centered images of the galaxy and
the PSF and a negligible error of the source flux $\mom{G}{0,0}$,
none of which is guaranteed for faint objects, the
deconvolution equations for the relevant second-order moments only mix
second-order moments. If furthermore $\Delta_G(2,s)=\Delta_P(2,s)$,
the ellipticity $\bchi$ (cf. \autoref{eq:ellipticity}) will remain
unbiased after deconvolution even though the moments themselves were
biased. The aforementioned condition holds if the radial profiles of
PSF and galaxy are similar within the weight function, in other
words, if the galaxy is small. 
This behavior can clearly be seen in the bottom panel of \autoref{fig:bias_weighting}, where the ellipticity estimate of the smaller elliptical galaxy (diamond markers) has sub-percent bias for $n_w\geq 2$ and $s\geq3$. The estimates for the larger galaxy are slightly higher because $|\Delta_G(2,s)|>|\Delta_P(2,s)|$, i.e. the deconvolution procedure overcompensates the PSF-induced change of the moments. However, sub-percent bias is achieved for $n_w\geq4$ and $s\geq5$.

For large galaxies, it might be advantageous to adjust the sizes of galaxy and PSF independently as this would render $\Delta_G(2,s_G)$
more comparable to $\Delta_P(2,s_P)$. However, we found employing a
common size $s$ for both objects to be more stable for the small and
noisy galaxy images typically encountered in weak-lensing
applications. We therefore adjust the size $s$ such as to allow an
optimal measurement of the deweighted PSF moments. Since the main
purpose of the weighting is the reduction of noise in the measured
moments, one could improve the presented scheme by increasing $s$ for
galaxies with larger surface brightness such as to reduce the bias when the data quality permits.

\subsection{Deweighting variance}
\label{sec:deweighting_variance}
Being unbiased in a noise-free situation does not suffice for a practical weak-lensing application as the image quality is strongly degraded by pixel noise. 
We therefore investigate now the noise properties of the deweighted and deconvolved moments. 

The variance of the weighted moments is given by
\begin{equation}
\label{eq:weighted_variance}
\sigma^2\bigl(\mom{I_w}{i,j}\bigr) = \sigma_n^2 \int d\mathbf{x}\ W^2(\mathbf{x})\, x_1^{2i} x_2^{2j}
\end{equation}
since the noise is uncorrelated and has a vanishing mean. It is
evident from \autoref{tab:deweighting} that the variance of the
deweighted moments increases with the number of contributing terms,
i.e. with $n_w$. Less obvious is the response under changes of
$s$. While each moment accumulates more noise with a wider weight
function, the prefactors of the deweighting correction terms is
proportional to $s^{-n_w}$ such that their impact is reduced for
larger $s$. 

\begin{figure}
\includegraphics[width=\linewidth]{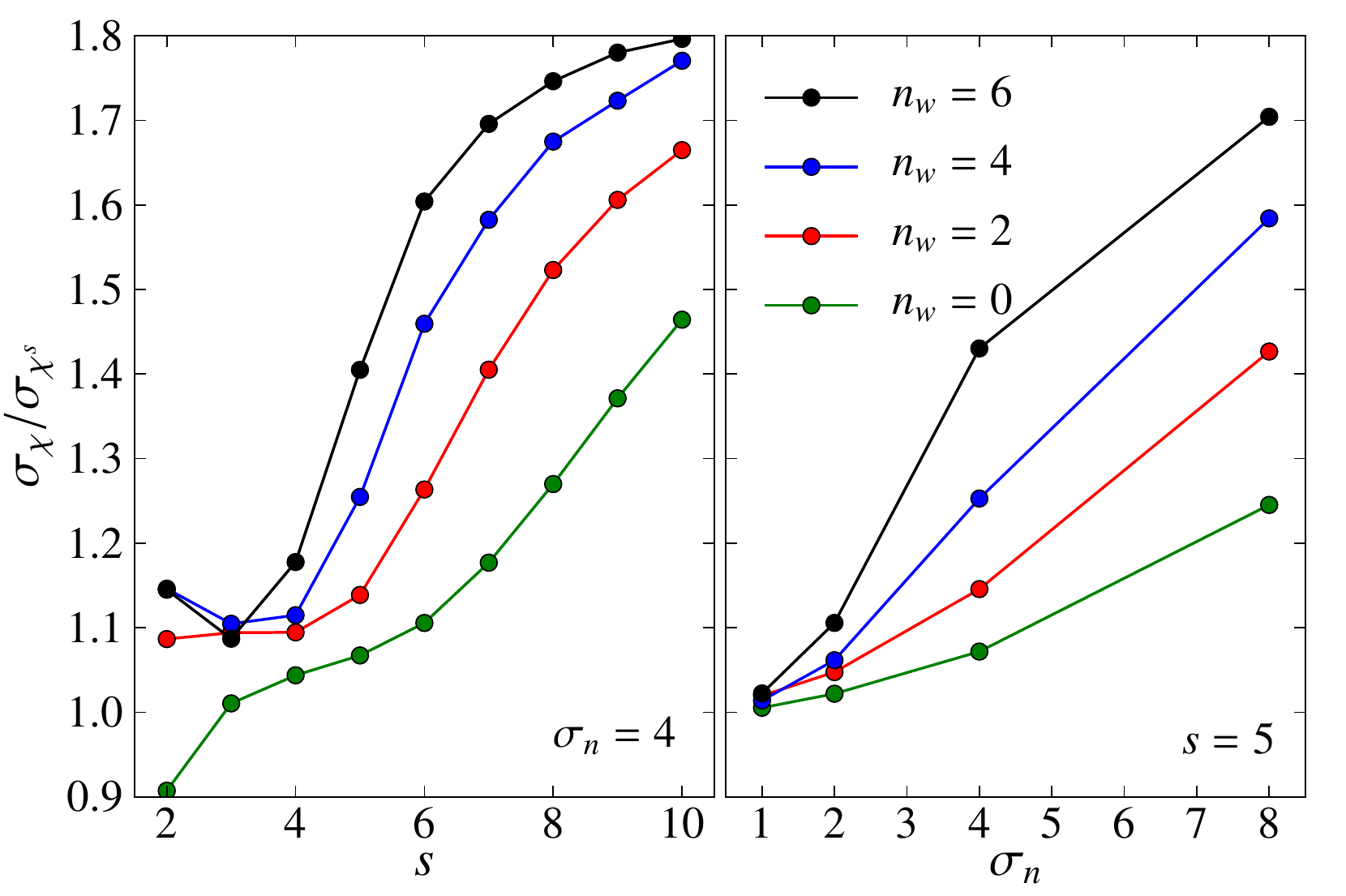}
\caption{Noise of the deweighted and deconvolved ellipticity estimate as a function of the weight function size $s$ (left) and of the standard deviation of the pixel noise $\sigma_n$ (right). The color and marker code is explained in \autoref{fig:bias_weighting}. The pixel noise is given in units of $10^{-3}$ for flux-normalized sources. $\sigma_n=8$ is close to the detection limit for this source model.}
\label{fig:noise_weighting}
\end{figure}

To quantitatively understand the impact of $n_w$ and $s$ in a fairly
realistic scenario we simulated 10,000 images of the galaxy models 1
and 2 from the last section. We drew their intrinsic ellipticities from
a Rayleigh distribution with $\sigma_{|\epsilon^s|} = 0.3$. Their flux
was fixed at unity, and the images were degraded by Gaussian pixel
noise with variance $\sigma_n^2$. We ran \deimos on each of
these image sets with a fixed scale $s$. The results are presented in
\autoref{fig:noise_weighting}, where we show the dispersion of the
measured $\bchi$ in units of the dispersion of $\bchi^s$. From the
left panel it becomes evident that the attempt of measuring unbiased
ellipticities (large $n_w$ or $s$) comes at the price of increased
noise in the estimates. Considering also \autoref{fig:bias_weighting},
we infer that in this bias-variance trade-off small values of $s$ and
large values of $n_w$ should be favored since this provides estimates with
high accuracy and a moderate amount of noise.

In the right panel of \autoref{fig:noise_weighting} we show the
estimator noise as function of the pixel
noise. \autoref{eq:weighted_variance} suggests that there should be a
linear relation between these two quantities, which is roughly
confirmed by the plot. Additional uncertainties in the moment
measurement -- caused by e.g. improper centroiding -- and the
non-linear combinations of second-order moments to form $\bchi$ lift
the actual estimator uncertainty beyond the linear prediction.

Even though the true errors of $\bchi$ may not exactly follow the linear theory, we will now exploit the fairly linear behavior to form error estimates. 
We can express the deweighting procedure as a matrix mapping,\begin{equation}
\mom{\mathbf{I}_{dw}}{} = \mathbfss{D}\cdot\mom{\mathbf{I}_w}{},
\end{equation}
where $\mom{\mathbf{I}_w}{}$ and $\mom{\mathbf{I}_{dw}}{}$ denote the vectors of all weighted and deweighted moments, and $\mathbfss{D}$ encodes the correction terms of \autoref{tab:deweighting}. The diagonal covariance matrix $\mathbfss{S}_w$ of the weighted moment variances given by \autoref{eq:weighted_variance} is related to the covariance matrix of the deweighted moments by
\begin{equation}
\mathbfss{S}_d = \mathbfss{D}\cdot \mathbfss{S}_w\cdot \mathbfss{D}^T,
\end{equation}
from which we can obtain the marginalized errors by
\begin{equation}
\label{eq:deweighted_variance}
\Bigl(\sigma^2\bigl(\mom{I_{dw}}{i,j}\bigr)\Bigl)^{-1} = \Bigl(\mathbfss{S}_d^{-1}\Bigr)_{k,k},
\end{equation}
where $k$ denotes the position of the moment $\mom{I_d}{i,j}$ in the vector $\mom{\mathbf{I}_{dw}}{}$. Under the assumptions mentioned above, also the deconvolution can be considered a linear operation, at least up to order 2, so that we can extend the error propagation even beyond this step: If we neglect errors in the PSF moments, the errors of the deconvolved moments (up to order 2) are identical to those of the deweighted ones. We can therefore estimate the errors of all quantities based on deconvolved moments directly from \autoref{eq:deweighted_variance}.

\section{Shear accuracy tests}
\label{sec:great08}

So far, we were concerned with the estimation of ellipticity. To test the ability of our new method to estimate the shear, we make use of
the reference simulations with realistic noise levels from the GREAT08 challenge
\citep{Bridle10.1}. As the shear values in these simulations are fairly low, we employ the linearized version of \autoref{eq:lensing_chi}, corrected by the shear responsivity of the source ensemble,
\begin{equation}
\tilde{\mathbf{g}} = \frac{\langle\bchi\rangle}{2-\sigma_\chi^2}
\end{equation}
\citep[e.g.][]{Massey07.1}, without any further weighting of individual galaxies, to translate \deimos ellipticity measures into shear estimates. The dispersion $\sigma_\chi^2$ is measured from the lensed and noisy galaxy images and hence only coarsely describes the intrinsic shape dispersion (cf. \autoref{fig:noise_weighting}). We are aware of this limitation and verified with additional simulations that it introduces sub-percent biases for the range of shears and pixel noise levels we expect from the GREAT08 images.

\begin{figure}
\includegraphics[width=\linewidth]{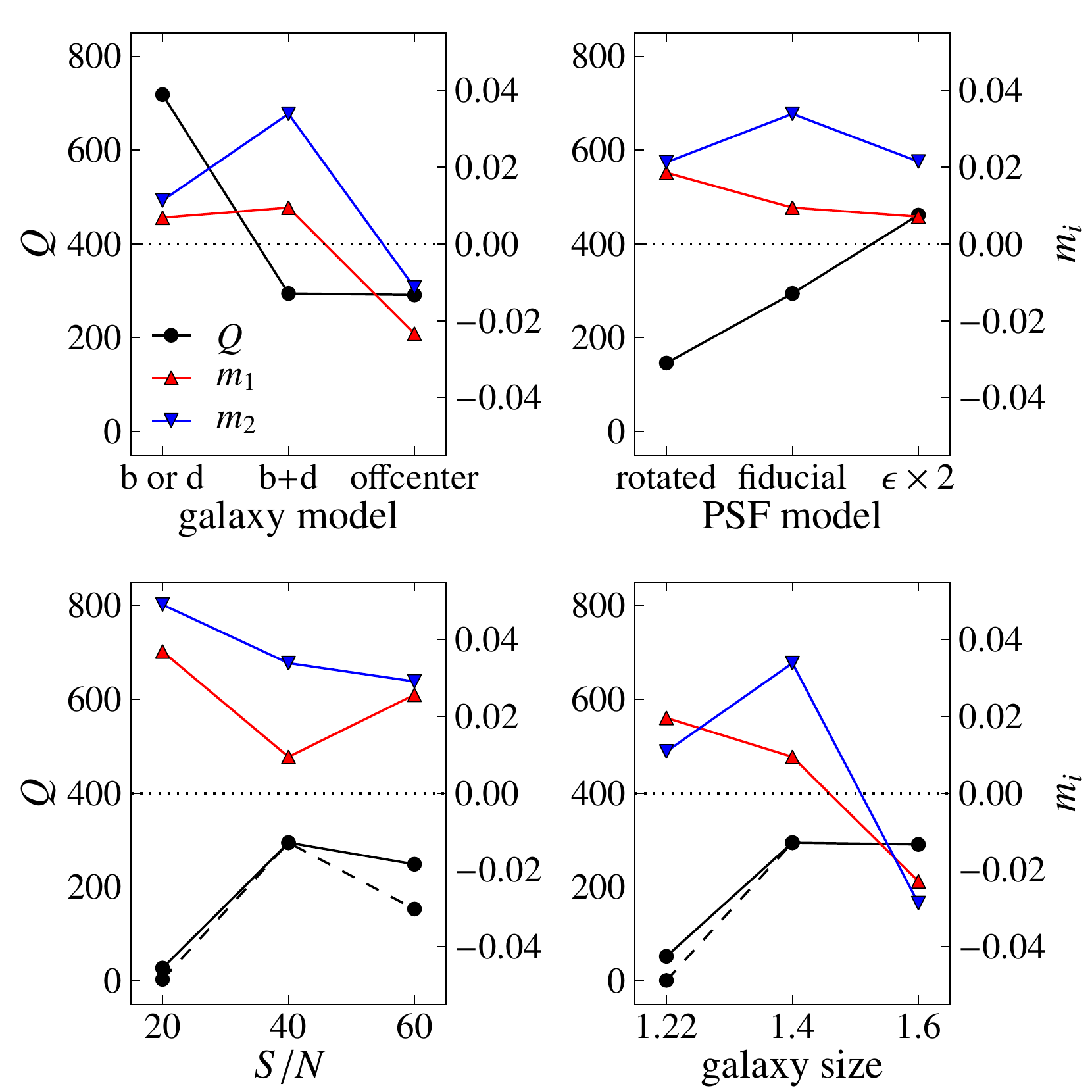}
\caption{GREAT08 $Q$ metric and multiplicative shear accuracy $m_i$
  for the nine different branches of the GREAT08 challenge data with
  realistic noise levels. In each panel, the scale on the left
  describes the values of $Q$ and the scale on the right the values of
$m_i$. The dotted line denotes $m_i=0$. The dashed
lines show the $Q$-value before we adjusted the weight-function matching
and deweighting parameters to the source characteristics of the branch.}
\label{fig:great08}
\end{figure}

We inferred the weight function size $s=4$ and the
correction order $n_w = 4$ from the optimal outcome for a set with known
shears. The actual GREAT08 challenge data comprises 9 different image sets, which differ in the shape of the PSF, the signal-to-noise ratio, the size,
and the model-type of galaxies. For each of these branches, there are 300 images
with different values of shear. We performed the \deimos
analysis of all images keeping the weighting parameters fixed to the
values inferred before. The results are shown in \autoref{fig:great08}
in terms of the GREAT08  quality metric $Q$ \citep[see eqs. 1\& 2
in][]{Bridle10.1} and of the multiplicative shear accuracy parameters $m_i$ obtained from a linear fit of the shear estimates $\tilde{g}_i$ to the true shear values $g_i$ \citep{Heymans06.1,Massey07.1},
\begin{equation}
\tilde{g}_i - g_i = m_i\, g_i + c_i.
\end{equation}

From \autoref{fig:great08} we clearly see the highly competitive
performance of \deimos with a typical $Q>200$ in all but two
branches. Single-component galaxy models yield a particularly large
$Q$-value, probably because the bulge-only models are the most
compact ones and thus favor the setting of a constant $s$ for PSF and
galaxies. In terms of $Q$, there is no change between the centered and
the off-centered double-component galaxy models, but both $m_i$ drop
for the off-centered ones. As such galaxy shapes have variable
ellipticity with radius and \deimos measures them with a fixed
weight function size, we interpret this as a small but noticeable
ellipticity-gradient bias \citep{Bernstein10.1}.

The response to changes in the PSF shape is a bit more worrisome and
requires explanation. The fiducial PSF had $\epsilon_1>\epsilon_2$,
and the opposite is true for the rotated one. From all panels of
\autoref{fig:great08} we can see that typically $|m_1| < |m_2|$. Such a
behavior has already been noted by \citet{Massey07.1}: Because a
square pixel appears larger in diagonal direction than along the pixel
edges, the moment $\mom{I}{11}$ and hence $\epsilon_2$ suffer more
strongly from the finite size of pixels. From the discussion in
\autoref{sec:deweighting_bias}, we
expect a certain amount of PSF-overcompensation for small weighting
function sizes. As the PSF shape is most strongly affected by
pixelation, the overcompensation boosts preferentially those galaxy
moments, which align with the semi-minor axis of the PSF. In general,
a larger PSF -- or a larger PSF ellipticity -- improves the shear estimates.
It is important to note, that, as in all other panels, the residual
additive term $c_i$ was negligible for all PSF models.

The response to changes in $S/N$ or galaxy size is more dramatic: Particularly the branches 7 (low $S/N$) and 9 (small galaxies) suffer from a considerable shear underestimation. This is not surprising as also most methods from \citet{Bridle10.1} showed their poorest performance in these two sets. 
Since the $Q$ metric strongly penalizes poor performance in single GREAT08 branches, the overall $Q=7.7$ for this initial analysis.

As this is the first application of \deimos to a weak-lensing test
case, we allowed ourselves to continue in a non-blind fashion in order
to work out how the \deimos estimates could be improved. Apparently,
problems arise when the galaxies are small or faint. The obvious
solution is to shrink the weight function size. As discussed in
section \autoref{sec:deweighting_variance}, improper centroiding plays
an increasing role in deteriorating shear estimates for fainter
galaxies. We therefore split the weight-function matching into two
parts: centroid determinations with a small weight function of size
$s_c$, and ellipticity determination with $s>s_c$. By choosing $s_c=
1.5$ and $s=2.5$, we could strongly improve the performance for
branches 7 and 9. Given the high $S/N$ of branch 6, we decided to
rerun these images with $n_w=6$, which yielded another considerable
improvement. With these modifications to the weight-function matching
and the deweighting parameters, \deimos estimates achieve $Q=112$,
similarly to {\scshape Lensfit} with $Q=119$, at a fraction of the runtime (0.015 seconds per GREAT08 galaxy). We emphasize that this
is a somewhat skewed comparison as we had full knowledge of the
simulation characteristics. However, the changes to the initial
analysis are modest and straightforward. In particular, they depend on
galactic size and magnitude only, and not on the true shears.

Given the bias-variance trade-off from the deweighting procedure, the
outcome of this section also clearly indicates that a simple \emph{one
  size fits all} approach is not sufficient to obtain highly accurate
shear estimates from \deimos. For a practical application, a scheme to
decide on $n_w$, $s_c$, and $s$ for each galaxy needs to be
incorporated. Such a scheme can easily be learned from a small set of
dedicated simulations, foremost because the \deimos results depend
only weakly on PSF and intrinsic galaxy shape. 

\section{Comparison to other methods}

Because of the measurement of image moments subject to a weighting
function, \deimos shares basic ideas and the computational
performance with the traditional KSB-approach \citep{Kaiser95.1}. In
contrast to it, \deimos does not attempt to estimate the
shear based on the ellipticity of single galaxies\footnote{This
  demands setting $\bchi^s=0$ in the non-linear
  \autoref{eq:lensing_chi}, which is only true on average but not
  individually.}, nor does it need to assume that the PSF can be
decomposed into an isotropic and an anisotropic part, which introduces
residual systematics into the shear estimation if the anisotropic part
is not small \citep{Kuijken99.1}. \deimos rather offers a
mathematically exact way of deconvolving the galaxy moments from any
PSF, thereby circumventing the problems known to affect KSB (see
\citet{Viola10.1} for a recent discussion of the KSB
shortcomings). Its only source of bias stems from the inevitably approximate
treatment of the weight function, which requires the measurement of
higher-order image moments. Since \deimos measures all moments with
the same weight function (instead of with increasingly narrower higher
derivatives of the weight function), these higher-order correction
terms suffer less from pixelation than those applied in KSB. However,
as we could see in \autoref{sec:great08}, pixelation affects the
\deimos measurements, and an analytic treatment of it is not obvious.

The treatment of the convolution with the PSF on the basis of moments is very close to the one known from shapelets \citep{Refregier03.2, Melchior09.1}. However, \deimos does not require the time-consuming modeling process of galaxy and PSF, and hence is not subject to problems related with insufficient modeling of sources, whose apparent shape is not well matched by a shapelet model of finite complexity \citep{Melchior10.1}.

In the RRG method \citep{Rhodes00.1}, the effect of the PSF convolution is also treated in moment space. Furthermore, an approximate relation between weighted and unweighted moments is employed, which renders this approach very similar to the one of \deimos. The former differs in the employment of the KSB-like anisotropy decomposition of the PSF shape.

As mentioned in \autoref{sec:weighting}, \deimos makes use of the same iterative algorithm as ELLIPTO \citep{Bernstein02.1} to define the centroid and ellipticity of the weight function. The latter additionally removes any PSF anisotropy by applying another convolution to render the stellar shapes circular, which is not necessary for \deimos.

The recently proposed FDNT method \citep{Bernstein10.1} deconvolves
the galaxy shape from the PSF in the Fourier domain, and then adjusts
centroid and ellipticity of the coordinate frame such that the
first-order moments and the components of the ellipticity -- formed
from second-order moments -- vanish in the new frame. FDNT restricts
the frequencies considered during the moment measurement to the
regime, which is not suppressed by PSF convolution. Because of the
shearing of the coordinate frame, additional frequencies need to be
excluded, whereby the allowed frequency regime further shrinks. This
leads to reduced significance of the shear estimates for galaxies with
larger ellipticities. Furthermore, FDNT requires complete knowledge of
the PSF shape. In contrast, \deimos does not need to filter the data,
it extracts the lensing-relevant information from the low-order
moments of the galaxy and PSF instead. These differing aspects
indicate that \deimos should be more robust against pixel noise. It
should also be possible to incorporate the correction for
ellipticity-gradient bias suggested by \citet{Bernstein10.1} in the
\deimos method.

\section{Conclusions}

For the presented work, we considered the most natural way of
describing the effects of gravitational lensing to be given by the
change of the multipole moments of background galaxies. We directly
estimate the lensed moments from the measured moments, which are
affected by PSF convolution and the application of a weighting
function. For the PSF convolution we derive an analytic relation
between the convolved and the unconvolved moments, which allows an
exact deconvolution and requires only the knowledge of PSF moments of
the same order as the galaxy moments to be corrected. The
weighting-induced changes of moments cannot be described analytically,
but for smooth weight functions a Taylor expansion yields
approximate correction terms involving higher-order moments.

We showed that the residual bias of the deweighted moments stemming
from an incomplete weighting correction is modest. Moreover, choosing
a weight function with matched ellipticities but same size for measuring
stellar and galactic moments yields ellipticity
estimates with very small bias even for rather small weighting
function sizes, which are required to reduce the impact of pixel noise
to a tolerable level. In this bias-variance 
trade-off, \deimos normally performs best with high correction orders $n_w$ at
small sizes $s$, but data with high significance may need a different
setup. The choice of these two parameters is the trickiest task for a
\deimos application, but can be easily addressed with a dedicated
simulation, which should resemble the size and brightness distribution
of sources to be expected in the actual data. Other properties of the
sources, like their ellipticity distribution or, more generally, their
intrinsic morphology, do not need to be considered as the measurement of
moments does neither imply nor require the knowledge of the true
source model.

There are certain restrictions of the method to bear in mind:
\begin{enumerate}
\item Setting $s$ to be the same for galaxies and the PSF works best
  for small galaxies, whose shape is dominated by the PSF shape. 
\item Changes of the shape at large radii would fall outside of
  the weight function and hence be ignored. When present in
  the PSF shape, this could lead to a residual PSF contamination, but
  can be cured by increasing the scale of the weight function at
  the expense of larger noise in the galaxy moments. When present in
  galactic shapes, the results become susceptible to
  ellipticity-gradient bias.
\item Direct measurement of the moments from the pixel values is
  inevitably affected by pixelation. For small, potentially
  undersampled shapes this leads to biased moment and ellipticity
  measures and acts more strongly in diagonal direction, i.e. on $\epsilon_2$.
\item The noise on the ellipticity estimates based on image moments is not
  Gaussian, nor does it propagate easily into the shear estimate. When
  dominant, it can create substantial biases of its own. 
\end{enumerate}
Only the first of these restrictions exclusively applies to \deimos,
the others are present in all non-parametric methods, which work
directly on the pixelated image. Model-based approaches could replace
the coarsely sampled moment measurements by ones obtained from the
smooth models.

Further work is required to choose the deweighting parameters, to account
for pixelation effects, and to address ellipticity-gradient bias
within the \deimos method. A C++ implementation of the method
described here can be requested from the authors. 

\section*{Acknowledgments}
PM is supported by the German Research Foundation (DFG) Priority
Programme 1177. MV is supported by the EU-RTN "DUEL" and by the IMPRS
for Astronomy and Cosmic Physics at the University Heidelberg.
BMS's work is supported by the DFG 
within the framework of the excellence initiative through the
Heidelberg Graduate School of Fundamental Physics.
\bibliography{../references}

\label{lastpage}
\end{document}